\documentclass[a4paper,11pt]{article}
\usepackage{jheppub}
\usepackage{array}
\usepackage{amsmath,amssymb,latexsym}
\usepackage{bm}
\usepackage[svgnames]{xcolor}

\usepackage{graphicx}
\usepackage{float}
\usepackage{subcaption}

\usepackage{mathtools}
\usepackage{physics}
\usepackage{braket}

\usepackage{natbib}
\usepackage{graphicx}

\usepackage{shuffle} 
\usepackage{slashed} 

\usepackage{multirow}
\usepackage{algorithm}
\usepackage{algpseudocode}
\usepackage{booktabs}

\allowdisplaybreaks

\usepackage{amsthm}
\newtheorem{lemma}{Lemma}
\renewenvironment{proof}[1][Proof]{\par\noindent\underline{\textbf{#1.}} \;}{\qed\par}

\preprint{CPTNP-2025-018}

\begin{document}

\title{Reconstructing Laurent expansion of rational functions using p-adic numbers}

\author{Tianya Xia,}
\author{Li Lin Yang}
\affiliation{Zhejiang Institute of Modern Physics, School of Physics, Zhejiang University, Hangzhou 310027, China}
\emailAdd{xiatianya@zju.edu.cn}
\emailAdd{yanglilin@zju.edu.cn}

\abstract{
We propose a novel method for reconstructing Laurent expansion of rational functions using $p$-adic numbers. By evaluating the rational functions in $p$-adic fields rather than finite fields, it is possible to probe the expansion coefficients simultaneously, enabling their reconstruction from a single set of evaluations. Compared with the reconstruction of the full expression, constructing the Laurent expansion to the first few orders significantly reduces the required computational resources. Our method can handle expansions with respect to more than one variables simultaneously. Among possible applications, we anticipate that our method can be used to simplify the integration-by-parts reduction of Feynman integrals in cutting-edge calculations.
}

\maketitle
                     
\newpage

\section{Introduction}

Algebraic manipulation and simplification of polynomial and rational expressions are basic operations encountered in almost all branches of mathematics and theoretical physics. There exist many general purpose computer algebra systems such as \texttt{Mathematica}, \texttt{Maple}, \texttt{Maxima}, \textit{et al.}, and dedicated programs such as \texttt{Fermat}~\cite{fermat}, \texttt{FORM}~\cite{Vermaseren:2000nd,Ruijl:2017dtg}, \texttt{GiNaC}~\cite{ginac}, \textit{et al.}, that can perform such operations. Nevertheless, manipulation of very large expressions can be extremely slow that often prohibits large scale applications of algebraic algorithms.

In high-energy physics, a typical situation where large expressions appear is in the perturbative calculation of scattering amplitudes and Feynman integrals. With the improvement of experimental precision at the Large Hadron Collider (LHC) and future high-energy colliders, it has become increasingly urgent to achieve high-precision theoretical predictions for important scattering processes. This involves higher perturbative orders for scattering amplitudes with multiple external legs. In these calculations, one needs to perform algebraic manipulations of large rational expressions to express the scattering amplitudes in terms of a set of master integrals (MIs), through the method of integration-by-parts (IBP) reduction \cite{Tkachov:1981wb,Chetyrkin:1981qh}. The tools for IBP reduction often employ \texttt{Fermat} for such manipulations, which proves to be more efficient than many general purpose computer algebra systems. However, this approach has reached its limit with the increasing complexity of the reduction tasks.

To overcome the difficulty, an important observation is that the final outcome of reduction is often much simpler than the intermediate expressions. This leads to the proposal that one may perform the reduction procedure numerically in a finite field $\mathbb{F}_p$, and only reconstruct the final expressions in the end \cite{Kant:2013vta}. This approach has been adopted in modern versions of reduction programs such as \texttt{Kira}~\cite{Klappert:2020nbg,Lange:2025fba}, \texttt{FIRE}~\cite{Smirnov:2019qkx}, and \texttt{LiteRed}~\cite{Lee:2012cn,Lee:2013mka} in combination with \texttt{FiniteFlow} \cite{Peraro:2019svx}. This has been combined with additional strategies of constructing reduced systems of IBP relations \cite{Wu:2023upw, Wu:2025aeg, Guan:2024byi, Song:2025pwy} to tackle cutting-edge problems.

However, even with these improvements, when the number of kinematic variables becomes larger, the final expressions may still be too complicated, such that the finite field reconstruction can become extremely slow. This has become one of the major bottlenecks towards calculating important physical quantities to higher precision.

To this end, it has been observed that the expressions expressed in the partial fraction form are often simpler than in the form with a common denominator, and dedicated programs \cite{Heller:2021qkz, Bendle:2021ueg} exist to perform such multivariate partial fraction for a known rational expression. This leads to the natural question: can we directly reconstruct the rational functions in the partial fraction form from numerical evaluations? There have been attempts towards this goal using $p$-adic numbers \cite{Wang:1981,DeLaurentis:2022otd,Chawdhry:2023yyx}, which face challenges when applying to large scale problems. For example, these methods require selecting a suitable set of integer evaluation points such that specific combinations of denominators are ``large'' in the sense of $p$-adic absolute value, which may not be always possible when the denominator structure is complicated. Furthermore, the prime number $p$ used in this method cannot be too large. As a result, one may need to employ a large number of distince $p$-adic fields, making the reconstruction process inefficient.

In this work, we propose a different way to utilize $p$-adic number in the reconstruction of rational expressions such as IBP reduction coefficients. This is based on the observation that we usually adopt dimension regularization in the calculation with the space-time dimension $d = 4 - 2 \epsilon$, and in practice, we are often only interested in the Laurent expansion of the results with respect to $\epsilon$, up to a specific order. Furthermore, in certain applications, we may also be interested in the asymptotic expansion of the expressions in certain kinematic limits, where one of the kinematic variables is much smaller than the others. This opens the possibility that one may directly reconstruct the coefficients in these series expansions, without knowing the full expression in advance. The main outcome of this work is that such reconstruction can be done using $p$-adic numbers, which can be significantly simpler than reconstructing the full results.

Before presenting our method, we note that \texttt{Ratracer}~\cite{Magerya:2022hvj} also supports the reconstruction of expanded rational functions, but in a different way. Starting from IBP identities, it first solves the equations and traces the solving steps. The program then expands the intermediate results to very high orders, and finally obtains the expanded functions. This is conceptually different from our approach, since we don't need to deal with (possibly large) intermediate expressions.

The remainder of this paper is organized as follows. In Section~\ref{section:intro_padic}, we briefly introduce the basic properties of $p$-adic numbers. Section~\ref{section:reconstruction_asy} presents the proposed method for reconstructing the Laurent expansion of rational functions. In Section~\ref{section:precision}, we briefly discuss some details about the computer representation of $p$-adic numbers. Section~\ref{section:denominator} discusses how the information about denominators in lower-order coefficients can be recycled to accelerate the reconstruction of higher-order ones. In Section~\ref{section:benchmark}, we demonstrate the efficiency of the proposed approach by comparing the number of probes and the time consumption per probe with the finite field reconstruction of the full expression. Finally, Section~\ref{section:summary} provides a brief summary of our findings.

\section{Basics of p-adic numbers}\label{section:intro_padic}

We begin by reviewing the essential mathematical background of $p$-adic numbers, which form a number field $\mathbb{Q}_p$. For a more detailed introduction, we refer the interested readers to \cite{Gouva2020}. Given a prime number $p$, a $p$-adic number $x$ can be defined as a series of the form
\begin{equation}
    \label{eq:p_adic_series}
    x = \sum_{i=v}^\infty a_i \, p^i \,,
\end{equation}
where each $a_i$ is an integer in the range $0 \leq a_i < p$, and $v$ is an (possibly negative) integer. If all $a_i$'s are zero, $x$ is the zero $p$-adic number, i.e., $x=0$. Otherwise, we require that $a_k > 0$, and $v$ is called the valuation of $x$, denoted as $v_p(x)$. In other words, $v_p(x)$ is the exponent of the leading term in the series definition of $x$. In general, the above series is not convergent in the usual sense. However, it is convergent in the sense of the $p$-adic absolute value, to be introduced later, where it will become clear that we should assign the valuation of $0$ to be $v_p(0) = +\infty$.

Before going into more formal aspects, let's get some intuition by looking at how integers and rational numbers can be represented as $p$-adic numbers. The simplest cases are positive integers. We are familiar with the concept of radix or base for representing numbers. For example, in base-$7$, the decimal integer $123$ is represented according to
\begin{equation}
    \label{eq:base7}
    123 = (234)_7 = 4 \times 7^0 + 3 \times 7^1 + 2 \times 7^2 \,,
\end{equation}
which is exactly in the form of a $p$-adic number. However, the difference between a number in base-$p$ and a $p$-adic number is already evident for negative integers. Take $-123$ as an example. In base-$7$, it is simply represented as $(-234)_7$. However, as a $p$-adic number, it is represented as the infinite series
\begin{equation}
    -123 = 3 \times 7^0 + 3 \times 7^1 + 4 \times 7^2 + \sum_{i=3}^\infty 6 \times 7^i \,.
\end{equation}
The result can be understood by noting that the sum of $123$ and $-123$ should be zero. From the above discussions, it is clear that the valuation $v_p(x)$ of a positive or negative integer $x$ is a non-negative integer. The addition, subtraction and multiplication operations of $p$-adic integers can be defined similarly as the operations on formal series.

We now turn to $p$-adic rational numbers. A $p$-adic rational number $r$ can be viewed as the division of a $p$-adic integer $n$ by another $p$-adic integer $m$ (assuming that $n$ and $m$ are coprime). Given the series representations of $n$ and $m$, that of $r$ can be constructed using $n = r \times m$. For example, we have
\begin{align}
    \frac{123}{7} &= 4 \times 7^{-1} + 3 \times 7^0 + 2 \times 7^1 \,, \nonumber
    \\
    \frac{123}{9} &= 2 \times 7^0 + 4 \times 7^1 + \sum_{i=2}^\infty 2 \times 7^i \,.
\end{align}
The first expression is easy to understand, while the second expression can be obtained from Eq.~\eqref{eq:base7} and $9 = 2 \times 7^0 + 1 \times 7^1$. It is clear that the valuation of a $p$-adic rational number is negative if the denominator $m$ has $p$ as a factor. In general, if $r = n/m$, one has $v_p(r) = v_p(n) - v_p(m)$. It is also evident that $r$ is usually an infinite series, unless $n$ is a positive integer and $m = p^l$ for some non-negative integer $l$.

The notion of absolute value for $p$-adic numbers is significantly different from that for real numbers. For a general field, an absolute value is a map from the field to real numbers satisfying the following properties: (1) $|x| \geq 0$ where the equality is achieved if and only if $x = 0$; (2) $|x y| = |x| |y|$; (3) $|x+y| \leq |x| + |y|$. There is a trivial possibility that satisfies all three conditions: $|x| = 0$ when $x=0$ and $|x|=1$ otherwise. This is the only possible absolute value for a finite field. If an absolute value satisfies the stronger condition $|x+y| \leq \max(|x|,|y|)$ for all $x$ and $y$, it is called a non-Archimedean absolute value. Otherwise it is an Archimedean one. The usual absolute value for real numbers is apparently Archimedean. On the other hand, the $p$-adic absolute value is non-Archimedean, and is defined by
\begin{equation}
    |x|_p = p^{-v_p(x)} \,.
\end{equation}
It is now clear that we need to assign $v_p(0) = +\infty$, and one may verify that the above definition satisfies the conditions for non-Archimedean absolute values. A counter-intuitive feature of this definition is that a larger rational number (larger in the usual sense) may have a smaller $p$-adic absolute value. For example, we have $|98|_p = 1/49$ and $|2/7|_p = 7$. 

The introduction of the absolute value promotes the $p$-adic field to a metric space, where the distance between two numbers $x$ and $y$ is given by $|x-y|_p$. Under this metric, it is easy to show that the series \eqref{eq:p_adic_series} is convergent (and absolutely convergent), since the higher-order terms in the series are actually ``smaller'' in the sense of absolute value. Another feature of the metric is that the set of $p$-adic rational numbers does not form a complete topological space, in the sense that a Cauchy sequence of rational numbers does not necessarily converge to a rational number. We are familiar with the concept that the set of real numbers $\mathbb{R}$ if the completion of the set of rational numbers $\mathbb{Q}$ equipped with the usual absolute value. Similarly, we say that $\mathbb{Q}_p$ is a completion of $\mathbb{Q}$ equipped with the $p$-adic metric, i.e., all Cauchy sequences of $p$-adic rational numbers can be written in the form \eqref{eq:p_adic_series}. An interesting behavior is that, for a $p$-adic rational number, the series \eqref{eq:p_adic_series} is either truncated at a finite order, or its coefficients will repeat themselves starting from some order. This is similar to the decimal representation of usual rational numbers.

We can now draw an analogy between a $p$-adic number $x \in \mathbb{Q}_p$ and a rational function expressed as a quotient of two polynomials:
\begin{equation}
    f(\bm{x},\lambda) = \frac{N(\bm{x},\lambda)}{D(\bm{x},\lambda)} \,,
\end{equation}
where $\bm{x}$ denotes a collection of variables, while $N(\bm{x},\lambda)$ and $D(\bm{x},\lambda)$ are polynomials in the variables $\bm{x}$ and $\lambda$. We would like to study the Laurent expansion of $f(\bm{x},\lambda)$ with respect to $\lambda$ near a point in the complex plane. Without loss of generality, we can set the point to be $\lambda = 0$.
\begin{equation}
    \label{eq:poly_expansion}
    f(\bm{x},\lambda) = \frac{N(\bm{x},\lambda)}{D(\bm{x},\lambda)} = \sum_{i = k}^\infty b_{i}(\bm{x}) \, \lambda^i \,,
\end{equation}
where each coefficient $b_i(\bm{x})$ is a polynomial in $\bm{x}$, and we require that $b_k(\bm{x})$ is not identically zero unless the function $f(\bm{x},\lambda)$ itself is zero.

Comparing Eq.~\eqref{eq:p_adic_series} and Eq.~\eqref{eq:poly_expansion}, we see that $p$-adic numbers provide a natural framework for studying the Laurent expansion of rational functions. In Eq.~\eqref{eq:p_adic_series}, each coefficient $a_i$ (vaguely) corresponds to an element in the finite field $\mathbb{F}_p$. This offers the possibility for reconstructing the coefficients $b_i(\bm{x})$ in Eq.~\eqref{eq:poly_expansion} using finite field techniques. We will outline the method of reconstruction in the next Section.

\section{Reconstruction of Laurent expansion of rational functions}
\label{section:reconstruction_asy}

In this section, we explore the possibility for reconstructing the Laurent expansion coefficients of a rational function $f(\bm{x},\lambda)$ with rational coefficients, using $p$-adic arithmetics. We will assume that we have a numerical algorithm to obtain $f(\bm{x},\lambda)$ as a $p$-adic number, with $\lambda = p$, and the other variables take value in $\mathbb{Z}_{>0}$. In many applications such as IBP reduction, this algorithm boils down to solving a system of linear equations. The goal is that, by repeating the numerical algorithm many times, we can reconstruct the analytic expressions of the expansion coefficients as rational functions. We will also discuss the possibility to perform simultaneous expansion in more than one variables.

\subsection{Finite field reconstruction of rational functions}

We begin by reviewing the reconstruction of rational functions using finite field techniques. For simplicity, we consider a univariate polynomial $f(x)$. The extensions to rational functions and to the multivariate cases are well-known in the literature \cite{Zippel:1979,Zippel:1990,Cuyt:2011,Monagan:2004,Ben-Or:1988,Kaltofen:1988,Kaltofen:1990,Kaltofen:1990_2,Kaltofen:2000,Kaltofen:2003,Javadi:2010}, and relevant algorithms have been implemented in the program packages \texttt{FiniteFlow} \cite{Peraro:2019svx} and \texttt{FireFly} \cite{Klappert:2019emp,Klappert:2020aqs}.

To proceed, we assume that the polynomial is of degree-$n$, where $n$ could be unknown before the reconstruction. We choose a sequence of distinct interpolation points $y_1,\cdots,y_{n+1}$, and write down an ansatz for the function in the form of Newton polynomials:
\begin{equation}
    f(x) = c_0 + \sum_{i=1}^n c_i \prod_{j=1}^i (x - y_j) \,.
\end{equation}
By evaluating $f(x)$ at these interpolation points in the finite field $\mathbb{F}_p$, we can obtain the coefficients $c_i$, also as an element in $\mathbb{F}_p$. In other words, what we obtain are actually $c_i$ modulo $p$. The maximal exponent $n$ can be determined when a couple of sequential coefficients are zero.

After reconstructing the function over $\mathbb{F}_p$, we need to promote the coefficients to rational numbers. Apparently, knowing $c_i$ modulo $p$ does not uniquely determine $c_i$ itself. Therefore, we usually need to introduce several different finite fields $\mathbb{F}_{p_i}$, and employ the Chinese Remainder Theorem (CRT) to determine the coefficients as rational numbers. In this respect, it is often preferred to choose as large prime numbers $p_i$ as possible. This can reduce the number of finite fields to be introduced, since the maximum integer that can be reliably reconstructed using the CRT is approximately $\prod_{i} p_i$. In practice, one often chooses the prime numbers to be slightly smaller than the maximal machine-size integer, which on a 64-bit system is $2^{63}-1$.

Since finite-field reconstruction is an indispensable ingredient in our method, we would also prefer large prime numbers for defining the $p$-adic field $\mathbb{Q}_p$. There is an additional benefit of using a large $p$: if a rational function $f(\bm{x},\lambda)$ does not have $\lambda$ as an overall factor, it is highly unlikely that evaluating the function in $\mathbb{Q}_p$ as $f(\bm{y},p)$ would lead to a $p$-adic number with an overall factor of $p$. In other words, the valuation of $f(\bm{y},p)$ precisely corresponds to the minimal exponent in the Laurent expansion of $f(\bm{x},\lambda)$. This largely avoids potential ambiguities and improves numerical stability in the reconstruction procedure.

As an illustrative example, consider a simple linear function
\begin{equation}
    f(x,\lambda) = b_0(x) + \lambda \,,
\end{equation}
where $b_0$ is a nonzero polynomial of $x$. For $f(y,p)$ to have an overall factor of $p$ with a randomly chosen $y \in \mathbb{Q}_p$, we will need $b_0(y) = 0 \pmod p$. The probability for this to happen becomes extremely low for large enough $p$. Similar considerations apply for higher degree polynomials and for rational functions. Therefore, when $p$ is sufficiently large, it is generally safe to assume that the $p$-adic expansion of $b_i(\bm{y})$ begins at $p^0$, and that of $f(\bm{y},p)$ begins at $p^k$ where $k$ is the minimal exponent in the Laurent expansion of $f(\bm{x},\lambda)$. This allows us to determine $k$ using merely one or two probes, and also allows us to reconstruct $b_i(\bm{x})$ using finite field techniques.

\subsection{Laurent expansion with respect to a single variable}

We consider a function which depends on variables $\bm{x}$ in addition to the variable $\lambda$ used for Laurent expansion:
\begin{equation}
    f(\bm{x},\lambda) = \sum_{i = k}^\infty b_{i}(\bm{x}) \, \lambda^i \,.
\end{equation}
By evaluating the function with $\lambda = p$ and $\bm{x} = \bm{y}$, where elements of $\bm{y}$ belong to $\mathbb{Z}_{>0}$, we obtain a $p$-adic number
\begin{equation}
    \label{eq:f_expansion}
    f(\bm{y},p) = \sum_{i = k}^\infty b_i(\bm{y}) \, p^i = \sum_{i = k}^\infty c_i(\bm{y}) \, p^i \,.
\end{equation}
Note that we can't directly assume $b_i(\bm{y}) = c_i(\bm{y})$, since $c_i(\bm{y})$ is an integer satisfying $0 \leq c_i(\bm{y}) < p$, while $b_i(\bm{y})$ is a $p$-adic number which itself admits the expansion
\begin{equation}
    \label{eq:b_expansion}
    b_{i}(\bm{y}) = \sum_{j=0}^\infty a_{i,j}(\bm{y}) \, p^{j} \,,
\end{equation}
where $a_{i,j}(\bm{y})$ are integers in the range $[0,p-1]$. Here, the minimal value of $j$ is $0$ for sufficiently large $p$, according to the discussions in the previous Section. Most importantly, this means that $a_{i,0}(\bm{y}) = b_i(\bm{y}) \pmod p$, and we can employ the finite field techniques to reconstruct $b_i(\bm{x})$ from the knowledge of $a_{i,0}(\bm{y})$.

Substituting Eq.~\eqref{eq:b_expansion} into Eq.~\eqref{eq:f_expansion}, and comparing the coefficients, we can extract the following relations
\begin{align}
    \label{eq:c_and_a}
    c_k(\bm{y}) &= a_{k,0}(\bm{y}) \,, \nonumber
    \\
    c_{k+1}(\bm{y}) &= a_{k,1}(\bm{y}) + a_{k+1,0}(\bm{y}) \pmod p \,, \nonumber
    \\
    c_{k+2}(\bm{y}) &= a_{k,2}(\bm{y}) + a_{k+1,1}(\bm{y}) + a_{k+2,0}(\bm{y}) + q_{k+1}(\bm{y}) \pmod p \,,
\end{align}
and so on, where $q_{k+1}(\bm{y})$ denotes the possible ``carry'' from the previous order. The reconstruction then proceeds in the following way. First, using the first equation, we obtain $a_{k,0}(\bm{y})$ from the value of $c_k(\bm{y})$. This allows us to reconstruct the analytic expression of $b_k(\bm{x})$ using finite field techniques. Once we have $b_k(\bm{x})$, we know the value of $a_{k,1}(\bm{y})$ for any $\bm{y}$, and we can then extract the value of $a_{k+1,0}(\bm{y})$ from the second equation in \eqref{eq:c_and_a}. This in turn allows us to reconstruct the rational function $b_{k+1}(\bm{x})$. This process can be performed iteratively for the next orders, until we reach the desired accuracy of the expansion, i.e., when $i = i_{\text{max}}$.

An important feature of the above reconstruction procedure is that the information about all desired coefficients $b_i(\bm{x})$ is contained in Eq.~\eqref{eq:f_expansion}. That is, we don't need to reevaluate the function for the construction at each order. This is the major benefit of using $p$-adic numbers. The required number of probes depends on the most complicated function in the list $\{b_k(\bm{x}),\cdots,b_{i_{\text{max}}}(\bm{x})\}$. Here, the complexity of a rational function mainly depends on the number of monomials in its numerator and denominator. Our method is most useful if this complexity is much lower than that of the original function $f(\bm{x},\lambda)$.

\subsection{Laurent expansion with respect to more than one variables}

Now let us consider the Laurent expansion of a rational function $f(\bm{x}, \lambda_1, \lambda_2)$ with respect to the two variables $\lambda_1$ and $\lambda_2$. The function can be expressed as a double series of the form
\begin{equation}
    \label{eq:two_var_expansion}
    f(\bm{x},\lambda_1,\lambda_2) = \sum_{i = k_1,\, j = k_2}^\infty b_{i,j}(\bm{x}) \, \lambda_1^{i} \, \lambda_2^{j} \,.
\end{equation}
The all possible values of $(i,j)$-pair form a lattice on the two-dimensional plane. In practice, we will truncate the series at some maximal values of $i$ and $j$, which we take as
\begin{equation}
    i_{\text{max}} = k_1 + h_1 - 1 \,, \quad j_{\text{max}} = k_2 + h_2 - 1 \,.
\end{equation} 
That is, there are $h_1$ distinct powers of $\lambda_1$, and $h_2$ distinct powers of $\lambda_2$. These values of $(i,j)$ form a subspace of the whole lattice, which we denote as
\begin{equation}
    \mathcal{D} = \{ (i,j) \, | \, k_1 \leq i \leq i_{\text{max}}, \, k_2 \leq j \leq j_{\text{max}} \} \,.
\end{equation}

In the $p$-adic reconstruction, we need to separate the information of each $b_{i,j}(x)$ into a unique power of $p$. For that purpose we choose $\lambda_1 = p^{n_1}$ and $\lambda_2 = p^{n_2}$ for some positive integers $n_1$ and $n_2$. The evaluation at $\bm{x} = \bm{y}$ then leads to 
\begin{equation}
    f(\bm{y},p^{n_1},p^{n_2}) = \sum_{i = k_1,\, j = k_2}^\infty b_{i,j}(\bm{y}) \, p^{n_1 i + n_2 j} \,.
    \label{eq:multivar}
\end{equation}
The integers $n_1$ and $n_2$ need to satisfy two conditions. Firstly, if $(i,j)$ and $(i',j')$ are two distinct pairs within $\mathcal{D}$, we require that $n_1 i + n_2 j \neq n_1 i' + n_2 j'$, or $n_1(i-i') \neq n_2(j'-j)$. Secondly, if the pair $(i',j')$ is outside $\mathcal{D}$, we require that no $(i,j)$ within $\mathcal{D}$ can satisfy $n_1(i-i') = n_2(j'-j)$. These two conditions ensure that the information of $b_{i,j}(\bm{y})$ for $(i,j)$ within $\mathcal{D}$ can be unambiguously extracted from the $p$-adic evaluation of the function.

As a simple choice, we take $n_1$ and $n_2$ to be coprime, which also satisfy $n_1 \geq h_2$ and $n_2 \geq h_1$. It is straightforward to verify that they satisfy the two conditions outlined above using the following lemma:
\begin{lemma}
Let positive integers $a, b \in \mathbb{Z}_{>0}$ be coprime. Consider the equation $n = a x + b y$ with integer $n \in [0, 2ab - a - b]$. If there exist a solution $(x,y)$ in the range $0 \leq x < b$ and $0 \leq y < a$, then it is the unique solution in the range $x \geq 0$ and $y \ge 0$.
\begin{proof}
    We prove by contradiction. Suppose that there exist two distinct solutions $(x_1,y_1)$ and $(x_2,y_2)$, with $0 \leq x_1 < b$, $0 \leq y_1 < a$, $x_2 \geq 0$ and $y_2 \geq 0$. We then have
    \begin{equation}
        n = ax_1 + by_1 = ax_2 + by_2 \,.
    \end{equation}
    Therefore
    \begin{equation}
        a(x_1-x_2) = b(y_2-y_1) \,.
    \end{equation}
    Since $ \gcd(a, b) = 1 $ and $(x_1,y_1) \neq (x_2,y_2)$, there must exist a non-zero integer $m$ such that
    \begin{equation}
        x_1 - x_2 = m b \,, \quad y_2 - y_1 = m a \,.
    \end{equation}
    If $m > 0$, we have $x_2 = x_1 - mb < 0$; if $m < 0$, we have $y_2 = y_1 + ma < 0$. This contradicts the assumption that both $x_2$ and $y_2$ are non-negative.
\end{proof}
\end{lemma}

The range of $n$ in the above lemma is constrained by the conditions $ 0 \leq x < b $ and $ 0 \leq y < a $, since
\begin{equation}
    n = ax + by \leq a(b - 1) + b(a - 1) = 2ab - a - b \,.
\end{equation}
Similar conclusions can be drawn from the lemma if the ranges of $x$ and $y$ are shifted by some constants $k_1$ and $k_2$. Therefore, by truncating the $p$-adic numbers in Eq.~\eqref{eq:multivar} at order $n_1 i_{\text{max}} + n_2 j_{\text{max}}$, we can reconstruct the relevant coefficient functions $b_{i,j}(\bm{x})$ using the method outlined previously.

\section{Precision of p-adic numbers in computer systems}
\label{section:precision}

In general, a $p$-adic number is represented as an infinite series \eqref{eq:p_adic_series}. However, a computer cannot handle infinitely long sequences, and we need to truncated the series to a given number of terms, which we refer to as the precision of the (representation of) $p$-adic numbers.

The above concept of precision is similar to the floating-point numbers in computer systems. However, there is an important difference between floating-point numbers and $p$-adic numbers. For floating-point numbers, arithmetic operations may suffer from so-called round-off errors, leading to potential loss of precision. While computer systems usually use the binary numbers, let's illustrate the point using decimal numbers for simplicity. Imagine that our system represent floating-point numbers with 4 significant decimal digits, and we want to compute $3 \sqrt{2}$. The computer may calculate it as $3 \times 1.414 = 4.242$. However, the true value (with 4 significant digits) should be $4.243$. The point here is that the ignored digits may accumulate giving rise to a carry, which is not accounted for.

Such round-off error is not a problem for $p$-adic numbers, since the ``carry'' from the last digit is added ``to the right'' which can be discarded, instead of ``to the left'' which should be kept. Let's consider an illustrative example. Suppose we have two $p$-adic numbers $x_1$ and $x_2$ that are truncated to the precision $h$:
\begin{align}
    x_1 = p^{v_1} \left( \sum_{i=0}^{h-1} a_{1,i} \, p^i + \mathcal{O}(p^h) \right) , \quad
    x_2 = p^{v_2} \left( \sum_{i=0}^{h-1} a_{2,i} \, p^i + \mathcal{O}(p^h) \right) ,
\end{align}
where we have extracted the valuation part such that the index $i$ starts from $0$. Without loss of generality, we assume that $v_1 \leq v_2$, or $\delta_v \equiv v_2 - v_1 \geq 0$. We now consider the sum $x_1+x_2$:
\begin{equation}
    x_1 + x_2 = p^{v_1} \left( \sum_{i=0}^{h-1} \left[ a_{1,i} + \theta(i-\delta_v) \, a_{2,i-\delta_v} \right] \, p^i + \mathcal{O}(p^h) \right) = p^{v_1} \left( \sum_{i=0}^{h-1} c_i \, p^i + \mathcal{O}(p^h) \right) ,
    \label{eq:addition}
\end{equation}
where $\theta(i)$ equals $1$ for $i \geq 0$, and equals $0$ otherwise. Here, each $c_i$ receives contribution from the corresponding square-bracket $\left[ a_{1,i} + \theta(i-\delta_v) \, a_{2,i-\delta_v} \right]$, with possible carry from the $p^{i-1}$ term. The last ``digit'' with $i=h-1$ may also give rise to a carry, which however should be added to the $p^h$ term. Therefore we see that all $h$ ``digits'' $c_0,\cdots,c_{h-1}$ are accurate and not affected by the truncation. Similar considerations apply for subtraction, multiplication and division.

Nevertheless, there is one situation that may lead to a precision loss. Suppose we have $v_1 = v_2 = v$ in Eq.~\eqref{eq:addition}, and it happens that $a_{1,0} + a_{2,0} = p$. In this case the valuation of $x_1+x_2$ becomes $v+1$, and we have
\begin{equation}
    x_1 + x_2 = p^{v+1} \left( \sum_{i=0}^{h-2} c_i \, p^i + \mathcal{O}(p^{h-1}) \right) .
\end{equation}
Hence, the precision of the result becomes $h-1$ instead of $h$. The probability for this to happen can be reduced by choose a large prime number $p$. When this is unavoidable, we can also simply increase the precision by one or two digits in the evaluation.

\section{Recycling information in the reconstruction of different orders}
\label{section:denominator}

In Section~\ref{section:reconstruction_asy}, we have emphasized that the reconstruction of different orders in the Laurent expansion utilizes the information from the same $p$-adic evaluations $f(\bm{y},p)$. This is the main advantage of our approach. In practice, the reconstruction is still performed order-by-order, and there is an additional kind of information that can be recycled. We explain it in this Section.

Consider a rational function $f(\bm{x},\lambda)$ in the form
\begin{equation}
    f(\bm{x},\lambda) = \frac{a_0 + a_1 \lambda + a_2 \lambda^2 + \mathcal{O}(\lambda^3)}{b_0 + b_1 \lambda + b_2 \lambda^2 + \mathcal{O}(\lambda^3)} \,,
\end{equation}
where $a_i$ and $b_i$ are polynomials of $\bm{x}$ (whose dependence is suppressed for simplicity). We assume that $a_0$ and $b_0$ are coprime polynomials, and therefore the series expansion with respect to $\lambda$ can be written as
\begin{align}
    \label{eq:rational_expansion_1}
    f(\bm{x},\lambda) = \frac{a_0}{b_0} + \frac{a_1 b_0 - a_0 b_1}{b_0^2} \lambda + \frac{a_2 b_0^2 - a_1 b_0 b_1 + a_0 b_1^2 - a_0 b_0 b_2}{b_0^3} \lambda^2
    + \mathcal{O}(\lambda^3) \, .
\end{align}
At this point, it is important to recall that the complexity of finite field reconstruction for a rational function depends crucially on the number of free parameters to be fixed, which in turn depends on the polynomial degrees of the numerator and the denominator. Therefore, knowing in advance some information about the numerator and the denominator can help to speed up the reconstruction by reducing the number of required probes. In the above example, once we finished reconstructing the $\lambda^0$ coefficient $a_0/b_0$, the information about $a_0$ and $b_0$ can be recycled into the reconstruction of higher-order terms. In particular, the denominators can be extracted out without the need to reconstruct again. In practice, this can significantly improve the reconstruction efficiency. Based on the above considerations, we can design a simple algorithm shown as Algorithm~\ref{alg:1}, to recycle the denominator information in the reconstruction.

\begin{algorithm}
    \caption{Naively recycling the information of known denominators}
    \label{alg:1}
    \begin{algorithmic}
        \State Set $F = 1$.
        \For{$\alpha$ in $[1,h]$}
            \State Numerically multiply $F$ with the $\alpha$-th order coefficient.
            \State Reconstruct the analytical expression $C_{\alpha}$ from the above numerical evaluations.
            \State Extract the denominator $D$ of $C_{\alpha}$.
            \State Return $C_{\alpha}/{F}$.
            \State Set $F = D$.
        \EndFor
    \end{algorithmic}
\end{algorithm}

However, things may not be that simple in certain cases. In the above we have assumed that $a_0$ and $b_0$ are coprime, which may not be true. Let's consider the following example
\begin{align}
    \label{eq:rational_expansion_2}
    f(\bm{x},\lambda) &= \frac{a_0 c_0 e_0 + a_1 e_0 \lambda + a_2 \lambda^2 + \mathcal{O}(\lambda^3)}{d_0 e_0\left(b_0 c_0 + b_1 \lambda + b_2 \lambda^2 + \mathcal{O}(\lambda^3)\right)} \nonumber
    \\
    &= \frac{a_0}{b_0 d_0} + \frac{a_1 b_0 - a_0 b_1}{b_0^2 c_0 d_0} \lambda + \frac{a_2 b_0^2 c_0 - a_1 b_0 b_1 e_0 + a_0 b_1^2 e_0 - a_0 b_0 b_2 c_0 e_0}{b_0^3 c_0^2 d_0 e_0} \lambda^2
    + \mathcal{O}(\lambda^3) \,.
\end{align}
In this case, the $\lambda^0$ terms of the numerator and the denominator share a common factor $c_0 e_0$, while the denominator has an overall factor $d_0 e_0$. As a result, $c_0$ and $e_0$ do not appear in the lowest order of the expansion, while $b_0$ and $d_0$ appear in a different pattern in the denominators of higher-order terms. Therefore, we must be more careful when designing the algorithm for recycling the lower-order information.

By examining Eq.~\eqref{eq:rational_expansion_2}, we can observe that there two categories of denominator factors. The first category contains overall factors such as $d_0$ and $e_0$, whose powers stay the same at each order. We collectively denote the first category as $\bm{F} = \{ F_i \}$. The second category includes $b_0$ and $c_0$, whose powers increase as the order increases. We collectively denote the second category as $\bm{G} = \{ G_i \}$. In the reconstruction, we need to distinguish these two kinds of factors, and also need to determine the minimal order when each of $F_i$ and $G_i$ first appears. For that purpose, we adopt the following approach. We multiply the target coefficient function with a rational ansatz and perform numerical probes, which can be used to analyze the degrees of the numerator and the denominator. The observed degrees can then be used to infer the structure of the denominator of the coefficient function. We summarize the approach as Algorithm~\ref{alg:2}. This recycling approach proves highly efficient for complex expressions, and is the default recycling algorithm to be used.

\begin{algorithm}
    \caption{Recycling the information about denominator factors.}
    \label{alg:2}
    \begin{algorithmic}
        \State Set $\bm{F}=\{\}$ and $\bm{G}=\{\}$.
        \State Set $\bm{a}=\{\}$.
        \For{$\alpha$ in $[1,h]$}
            \State Numerically multiply $\prod_i F_i \prod_j G_j^{a_j}$ with the $\alpha$-th order coefficient.
            \State Reconstruct the analytical expression $C_{\alpha}$ from the above numerical evaluations.
            \State Extract the denominator $D$ of $C_{\alpha}$.
            \If{$D \neq 1$}
                \State Factorize $D$ into $\prod_{k=1}^{m} D_k^{n_k}$.
                \State Probe the degree $d_{N}$ of numerator and $d_{D}$ of the denominator for the $(\alpha+1)$-th order coefficient.
                \State Set $F_{\alpha+1}=1$ and $G_{\alpha+1}=1$.
                    \For{$\beta$ in $[1,m]$}
                        \State Set $d_{\beta}$ to the degree of $D_{\beta}$.
                        \State Numerically multiply $D_{\beta}^{2 n_{\beta}}$ with the $(\alpha+1)$-th order coefficient.
                        \State Probe the degree $d'_{N}$ of numerator and $d'_{D}$ of the denominator for the above result.
                        \If{$d'_{N} > d_{N}$}
                            \State $F_{\alpha+1} = F_{\alpha+1} \times D_{\beta}^{(d'_{N}-d_{N})/d_{\beta}}$.
                            \State $G_{\alpha+1} = G_{\alpha+1} \times D_{\beta}^{n_{\beta}-(d'_{N}-d_{N})/d_{\beta}}$.
                        \ElsIf{$d'_{N} = d_{N}$}
                            \State $G_{\alpha+1} = G_{\alpha+1} \times D_{\beta}^{n_{\beta}}$.
                        \EndIf
                    \EndFor
                \If{$F_{\alpha+1} \neq 1$ and $G_{\alpha+1} \neq 1$}
                    \State Append $F_{\alpha+1}$ to $\bm{F}$ and append $G_{\alpha+1}$ to $\bm{G}$.
                    \State Append $1$ to $\bm{a}$ and then let $\bm{a} = \bm{a} + \bm{1}$.
                \EndIf  
            \EndIf
            \State Return $\frac{C_{\alpha}}{\prod_i F_i \prod_j G_j^{a_j}}$.
        \EndFor
    \end{algorithmic}
\end{algorithm}

\section{Benchmarks in the reconstruction of IBP coefficients}
\label{section:benchmark}

As already mentioned in the Introduction, one of the main use case of our method is the reconstruction of IBP reduction coefficients. These coefficients are functions of the momentum invariants, the internal masses, as well as the dimensional regulator $d=4-2\epsilon$. In practice, we are usually interested in the expansion of Feynman integrals as Laurent series in $\epsilon$. Therefore, it is enough to have the reduction coefficients as Laurent series in $\epsilon$ as well. Furthermore, it is often useful to study the scattering amplitudes in certain kinematics limits. For that purpose, the relevant Feynman integrals and their reduction coefficients can also be expanded in those limits as well.

As a concrete example, we consider a particular reduction coefficient which appears in the calculation of a three-body massive form factor \cite{Guo:2025dlm}. The expression involves five variables: $\epsilon$, $m^2$, $s_{12}$, $s_{23}$ and $s_{123}$. The size of the expression stored in computer is about 400~KB. We are interested in its Laurent expansion with respect to $\epsilon$ or $m^2$. In Table~\ref{tab:391_eps}, we list the number of probes required for the reconstruction of the expansion coefficients in $\epsilon$. We show both the numbers for the reconstruction with or without recycling the information about the denominator factors. Similarly, Table~\ref{tab:391_m2} presents the corresponding numbers for the expansion in $m^2$. As a comparison, the full reconstruction of the expression in finite fields requires $313483$ probes. As one can see, comparing with the full expression, reconstructing the expansion coefficients requires orders-of-magnitude smaller numbers of probes. This is particularly true at low orders and when the denominator recycling is employed. Therefore, our method is highly efficient in complicated reduction tasks.

\begin{table}[t!]
    \centering
    \begin{tabular}{lll}
        \hline
        Order in $\epsilon$ & Without recycling & With recycling \\ \hline
        0 & 1334 & 1334 \\
        1 & 18853 & 5313 \\
        2 & 140124 & 5771 \\
        3 & 389795 & 13653 \\
        \hline
    \end{tabular}
    \caption{Number of probes required for the reconstruction of $\epsilon$-expansion coefficients for a 400~KB expression. As a comparison, the full reconstruction of the unexpanded expression requires $313483$ probes.}
    \label{tab:391_eps}
\end{table}

\begin{table}[t!]
    \centering
    \begin{tabular}{lll}
        \hline
        Order in $m^2$ & Without recycling & With recycling \\ \hline
        0 & 277 & 277 \\
        1 & 792 & 572 \\
        2 & 1760 & 1214 \\
        3 & 8749 & 4566 \\
        \hline
    \end{tabular}
    \caption{Number of probes required for the reconstruction of $m^2$-expansion coefficients for a 400~KB expression. As a comparison, the full reconstruction of the unexpanded expression requires $313483$ probes.}
    \label{tab:391_m2}
\end{table}

To understand the efficiency provided by the expansion, it is helpful to compare the maximal powers of the variables appearing in the denominators, both of the full expression and of the expansion coefficients (with the known denominator factors stripped). We show these information in Table~\ref{tab:391_eps_degree} for the $\epsilon$-expansion and Table~\ref{tab:391_mt2_degree} for the $m^2$-expansion. As one can see, the recycling of the denominators provides amazing simplification, particularly in the case of $\epsilon$-expansion. This explains the significant decrease of the number of probes observed in Table~\ref{tab:391_eps}. 

\begin{table}[t!]
    \centering
    \begin{tabular}{lcc}
        \hline
        Order in $\epsilon$ & Without recycling & With recycling \\ \hline
        0 & 12 & 12 \\
        1 & 22 & 6 \\
        2 & 32 & 0 \\
        3 & 41 & 0 \\
        \hline
    \end{tabular}
    \caption{Degrees of denominators in the reconstruction of $\epsilon$-expansion coefficients for a 400~KB expression. For reference, the degree of the denominator in the unexpanded expression is $31$.}
    \label{tab:391_eps_degree}
\end{table}

\begin{table}[t!]
    \centering
    \begin{tabular}{lcc}
        \hline
        Order in $m^2$ & Without recycling & With recycling \\ \hline
        0 & 5 & 5 \\
        1 & 9 & 3 \\
        2 & 13 & 2 \\
        3 & 20 & 4 \\
        \hline
    \end{tabular}
    \caption{Degrees of denominators in the reconstruction of $m^2$-expansion coefficients for a 400~KB expression. For reference, the degree of the denominator in the unexpanded expression is $31$.}
    \label{tab:391_mt2_degree}
\end{table}

At this point, one may raise a question. Even if the number of probes are significantly reduced, the computation in $p$-adic fields is, in principle, more complicated than the computation in finite fields. Will this computational cost overwhelms the benefits of the expansion? To answer this question, it suffices to compare straightforwardly the time consumption for each probe in both cases. We build an in-house \texttt{C++} program utilize the \texttt{FLINT} library for both computation in $p$-adic fields and finite fields. This can be combined with the \texttt{FireFly} library for rational reconstruction.

\begin{figure}[t!]
    \centering
    \includegraphics[width=0.7\textwidth]{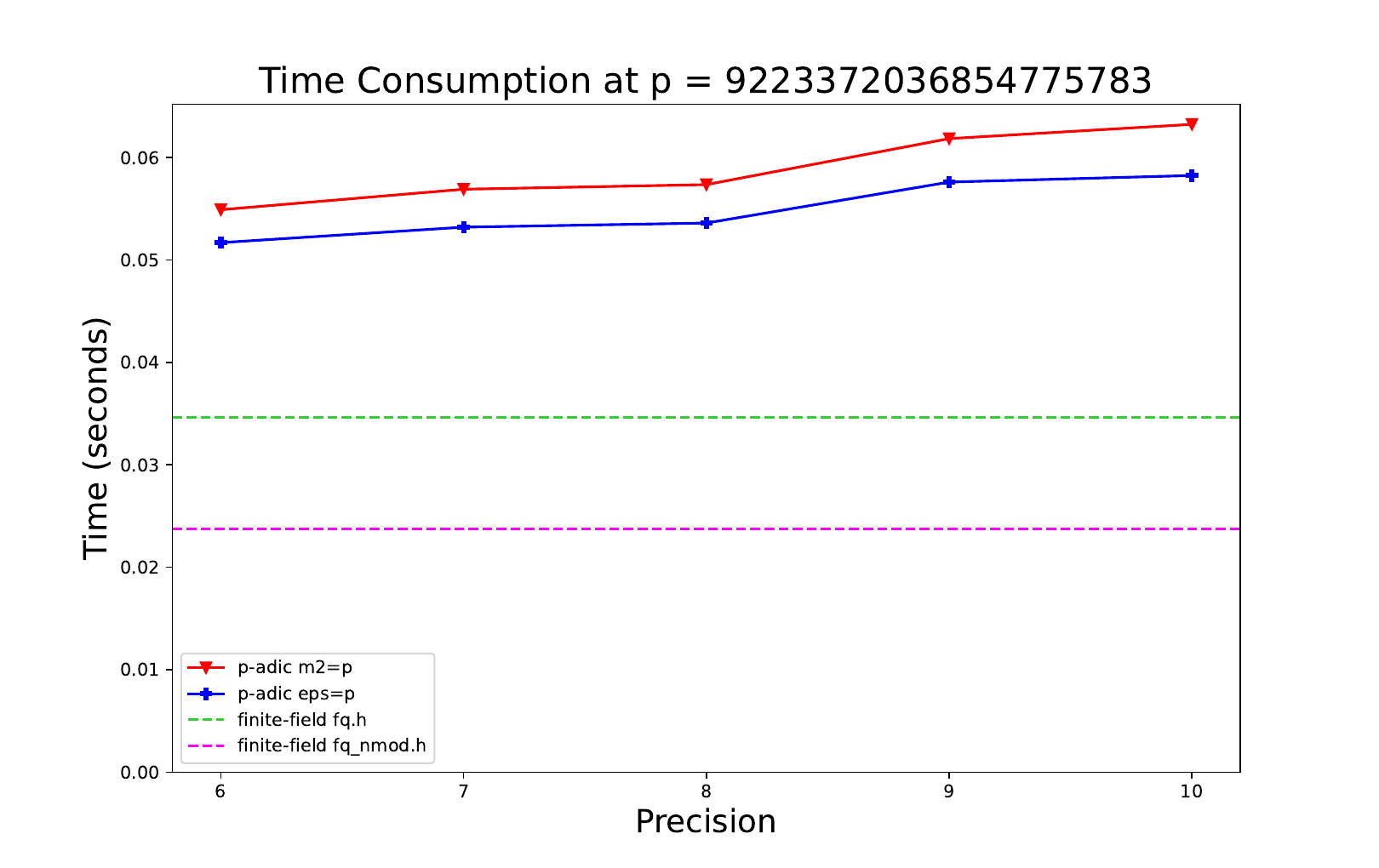}
    \\
    \includegraphics[width=0.7\textwidth]{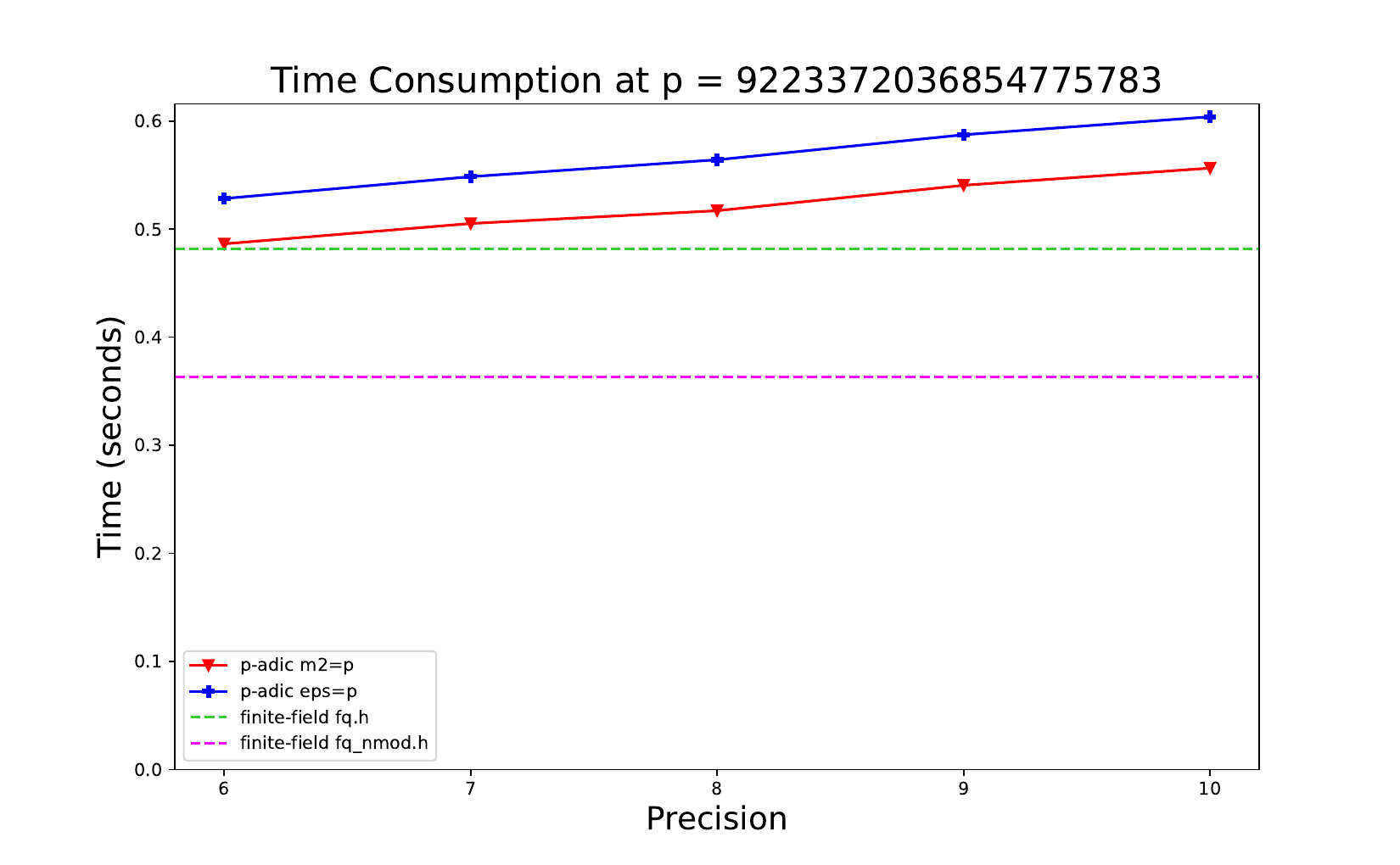}
    \caption{The time consumption per probe in the finite field and in the $p$-adic field. The dashed lines show the time consumption in the finite field using two different interfaces \texttt{fq.h} and \texttt{fq\_nmod.h} in \texttt{FLINT}. The solid lines represent the time consumption in the $p$-adic field as a function of the precision, with different variables set to $p$. The upper and lower plots correspond to the evaluation of a 400~KB expression and a 1.6~MB expression, respectively}
    \label{fig:compare_with_finite_field}
\end{figure}

In Figure~\ref{fig:compare_with_finite_field}, we show the time consumption per probe for evaluating two large expressions. The first is the 400~KB expression mentioned earlier, and the second is a 1.6~MB expression from the calculation in \cite{Guo:2025dlm}. We choose the prime number to be $p=9223372036854775783$, which is close to $2^{63}-1$. For the $p$-adic field, we choose several different settings for the precision, which are enough for the expansion in $\epsilon$ to the fifth order, and in $m^2$ to the second order. The resulting time consumption is show as the solid lines. Each solid line corresponds to setting one of the variables to $p$, which is relevant for the Laurent expansion with respect to this particular variable. For the finite field, the time consumption is show as the dashed horizontal lines, which is independent of the precision chosen for the $p$-adic representation. For the finite field, we have used two different interfaces \texttt{fq.h} and \texttt{fq\_nmod.h} in \texttt{FLINT}, and we observe that the latter is slightly faster than the former. The reason is that the latter interface only accept machine-size prime numbers smaller than $2^{63}-1$, while the former could be used when even bigger prime numbers are required in the reconstruction of more complicated expressions.

From Figure~\ref{fig:compare_with_finite_field}, one can see that, while the evaluation in the $p$-adic field is in general a bit slower than that in the finite field, the difference is not so big. The time consumption per probe in the $p$-adic field roughly stays around or below a factor of two with respect to that in the finite field. In view of the significant reduction of the number of probes, this slight increase of the time per probe does not affect the efficiency provided by our method. On the other hand, we emphasize that the numbers quoted here only provide a crude comparison. In physical applications, the algorithms for numerical evaluation can be quite different. In IBP reduction, e.g., the algorithm is usually Gaussian elimination for solving linear equations. The time consumption per probe should then be investigated case-by-case. However, since these algorithms mostly involve basic arithmetics, we believe that the main conclusion should be unchanged.

In the above, we have only considered the expansion with respect to either $\epsilon$ or $m^2$. As discussed previously, we may perform a simultaneous expansion in both $\epsilon$ and $m^2$ with our approach. Conceptually, the structure of the coefficients in the double expansion should be even simpler than those in the single expansions. It can therefore be expected that the required number of probes should be smaller. On the other hand, the required precision in the $p$-adic expansion needs to be larger. However, as can be observed from Figure~\ref{fig:compare_with_finite_field}, the increase of the time consumption per probe with respect to the precision is mild. As a result, we expect that the reconstruction of the double expansion should be more efficient than that of the single expansion.

\begin{table}[t!]
    \centering
    \begin{tabular}{lll}
        \hline
        Order in $\epsilon$ & Recycling Alg.~1 & Recycling Alg.~2 \\ \hline
        0 & 1334 & 1334 \\
        1 & 6439 & 5313 \\
        2 & 56107 & 5771 \\
        3 & 165503 & 13653 \\
        \hline
    \end{tabular}
    \caption{Number of probes required for the reconstruction of $\epsilon$-expansion coefficients for a 400~KB expression with two recycling algorithms.}
    \label{tab:391_eps_alg2}
\end{table}

\begin{table}[t!]
    \centering
    \begin{tabular}{lcc}
        \hline
        Order in $\epsilon$ & Recycling Alg.~1 & Recycling Alg.~2 \\ \hline
        0 & 12 & 12 \\
        1 & 10 & 6 \\
        2 & 10 & 0 \\
        3 & 9 & 0 \\
        \hline
    \end{tabular}
    \caption{Degrees of denominators in the reconstruction of $\epsilon$-expansion coefficients for a 400~KB expression with two recycling algorithms}
    \label{tab:391_eps_degree_alg2}
\end{table}

Finally, it might be helpful to compare the performance of the two recycling algorithms. We show a comparison of the required number of probes in Table~\ref{tab:391_eps_alg2}. We see that Algorithm~\ref{alg:2} is significantly better than Algorithm~\ref{alg:1}, especially at higher orders. This can be explained by the degrees of the denominators that need to be reconstructed, as shown in Table~\ref{tab:391_eps_degree_alg2}. On the other hand, Algorithm~\ref{alg:2} may still not be the most optimal one. From the structure of Eq.~\eqref{eq:rational_expansion_2}, one can see that the numerators at higher-orders also involve ingredients of lower-order coefficients. We expect that these information can also be recycled to further improve the efficiency of the reconstruction.

\section{Summary}\label{section:summary}

In this work, we propose a novel method for reconstructing Laurent expansion of rational functions from numerical evaluations in $p$-adic fields. Our method can be applied if there exists an algorithm using $p$-adic numbers for obtaining numerical values of the unknown rational functions. In particular, this applies to solutions of linear equations, and a typical situation in high-energy physics is the IBP reduction of Feynman integrals and scattering amplitudes.

The key point in our method is that the analytical expressions of the expansion coefficients are usually much simpler than the full rational function, and the coefficients at different orders can be probed simultaneously by evaluating in $p$-adic fields. Combining with a suitable recycling algorithm to share information across order, we have demonstrated that the number of probes required to reconstruct the first few orders in the expansion can be significantly reduced, compared with the full reconstruction of the rational function. Furthermore, we show that arithmetics in $p$-adic fields is not much slower than in finite fields. Taken together, our approach provides a highly efficient method to obtain the Laurent expansion of very complicated rational functions. We anticipate that our method can be used to simplify the IBP reduction of Feynman integrals in cutting-edge calculations.

\begin{acknowledgments}
This work was supported in part by the National Natural Science Foundation of China under Grant No. 12375097, 12347103, and the Fundamental Research Funds for the Central Universities.
\end{acknowledgments}

\bibliographystyle{JHEP}
\bibliography{references_inspire.bib}

\end{document}